\documentstyle[preprint,eqsecnum,aps]{revtex}
\begin{document}
\draft
\preprint{IFUM 452/FT, December 1993}
\title{A consistent derivation of the quark-antiquark\\and
three-quark potentials\\in a Wilson loop context}

\author{N. Brambilla, P. Consoli and G. M. Prosperi}

\address{
Dipartimento di Fisica dell'Universit\`{a} -- Milano\\
INFN, Sezione di Milano -- Via Celoria 16, 20133 Milano
}

\maketitle
\begin{abstract}
In this paper we give a new derivation of the quark-antiquark
potential in the Wilson loop context. This makes more explicit the
approximations involved and enables an immediate extension to the
three-quark case. In the $q\overline{q}$ case we find the same
semirelativistic potential obtained in preceding papers but for a
question of ordering. In the $3q$ case we
find a spin dependent potential identical to that already derived in
the literature from the ad hoc and
non correct assumption of scalar confinement. Furthermore
we obtain  the correct form of the spin independent  potential
up to the $1/m^2$ order.
\end{abstract}


\section{Introduction}

The aim of this paper is twofold.

First we give a simplified
derivation of the quark-antiquark potential in the context of the so
called Wilson loop approach [1] in which the basic
assumptions, the conditions for the validity of a potential
description and the
relation with the flux tube model [2] can be better appreciated.

Secondly we show how the procedure can be extended to the three-quark
system [3] obtaining consistently   not only the static
part (stat) of the potential but also the spin dependent (sd) and the
velocity dependent (vd) ones at the $1/m^2$ order.

For what concerns the $q\overline{q}$ potential the result is identical
to that reported in [4,5]  (see [6] for the spin dependent potential)
except for a problem of ordering of
minor phenomenological interest:
\begin{equation}
V^{q\overline{q}} = V^{q\overline{q}}_{{\rm stat}} +
V^{q\overline{q}}_{{\rm sd}} +
V^{q\overline{q}}_{{\rm vd}}
\label{eq:vdue}
\end{equation}
where
\begin{equation}
V^{q\overline{q}}_{{\rm stat}} = -  \frac{4}{3}
 \frac{{\alpha}_s}{r} + \sigma r \> ,
\end{equation}
\begin{eqnarray}
V^{q\overline{q}}_{{\rm sd}} &=&
\frac{1}{8} \left( \frac{1}{m_1^2}
 + \frac{1}{m_2^2} \right)
\nabla^2 \left( - \frac{4}{3} \frac{\alpha_s}{r} + \sigma r \right)
+
\nonumber\\
{} &+&  \frac{1}{2} \left(
 \frac{4}{3} \frac{\alpha_s}{r^3} -
\frac{\sigma}{r} \right) \left[ \frac{1}{m_1^2} {\bf S}_1 \cdot
( {\bf r} \times {\bf p}_1 ) - \frac{1}{m_2^2} {\bf S}_2 \cdot
( {\bf r} \times {\bf p}_2 ) \right] +
\nonumber\\
{} &+& \frac{1}{m_1m_2} \frac{4}{3} \frac{\alpha_s}{r^3} [ {\bf S}_2
\cdot ( {\bf r} \times {\bf p}_1 ) - {\bf S}_1 \cdot ( {\bf r}
\times {\bf p}_2 )] +
\nonumber\\
&+& \frac{1}{m_1m_2} \frac{4}{3} \alpha_s \left\{ \frac{1}{r^3}
\left[ \frac{3}{r^2} ({\bf S}_1 \cdot {\bf r})({\bf S}_2 \cdot
{\bf r}) - {\bf S}_1 \cdot {\bf S}_2 \right] +
\frac{8\pi}{3} \delta^3({\bf r}) {\bf S}_1 \cdot {\bf S}_2 \right\}
\> ,
\end{eqnarray}
\begin{eqnarray}
V^{q\overline{q}}_{{\rm vd}} & = &
  \frac{1}{2m_1m_2} \left\{
 \frac{4}{3} \frac{{\alpha}_s}{r}
(\delta^{hk} + \hat{r}^h \hat{r}^k) p_1^h p_2^k \right\}_{{\rm W}} -
\nonumber\\
{} & - & \sum_{j=1}^2 \frac{1}{6m_j^2} \{ \sigma \, r \,
 {\bf p}_{j{\rm T}}^2  \}_{{\rm W}} -
\frac{1}{6m_1m_2} \{ \sigma \, r \,
{\bf p}_{1{\rm T}} \cdot {\bf p}_{2{\rm T}} \}_{{\rm W}} \> .
\end{eqnarray}
Obviously in Eqs. (1.2)-(1.4)
${\bf r}={\bf z}_1 -{\bf z}_2$  denotes the relative position
 of the quark and the antiquark and
${\bf p}_{j {\rm T}}$ the transversal momentum of the particle $j$,
$ p_{j{\rm T}}^h =
 ( \delta^{hk} -\hat{r}^h \hat{r}^k) p^k_j $
where $\hat{{\bf r}} =({\bf r}/r)$; the symbol $\{ \,\,\}_{\rm
W}$ stands for the Weyl ordering
prescription  among momentum  and position variables (see Sec. IV).
Furthermore  in comparison
with [5] the terms in the zero point  energy $C$ have been
omitted, since they should  be reabsorbed  in a redefinition
of the masses in a full relativistic treatment.

For the $3q$ potential the result is
\begin{equation}
V^{3q} = V^{3q}_{{\rm stat}} + V^{3q}_{{\rm sd}} +
V^{3q}_{{\rm vd}}
\end{equation}
with
\begin{equation}
V^{3q}_{{\rm stat}} =
\sum_{j<l} \left( -\frac{2}{3}
\frac{\alpha_s}{r_{jl}} \right) + \sigma (r_1+r_2+r_3) \> ,
\end{equation}
\begin{eqnarray}
V^{3q}_{{\rm sd}} & = &
\frac{1}{8m_1^2} \nabla^2_{(1)} \left( -\frac{2}{3}
\frac{\alpha_s}{r_{12}} -\frac{2}{3} \frac{\alpha_s}{r_{31}} +
\sigma r_1 \right) +
\nonumber\\
{}&+& \left\{ \frac{1}{2m_1^2} {\bf S}_1 \cdot \left[
({\bf r}_{12} \times {\bf p}_1) \left( \frac{2}{3}
\frac{\alpha_s}{r_{12}^3} \right) + ({\bf r}_{31} \times
{\bf p}_1) \left( -\frac{2}{3} \frac{\alpha_s}{r_{31}^3}
\right) - \frac{\sigma}{r_1} ({\bf r}_1 \times {\bf p}_1)
\right]  \right. +
\nonumber\\
{} &+& \left. \frac{1}{m_1m_2} {\bf S}_1 \cdot ({\bf r}_{12}
\times {\bf p}_2) \left( -\frac{2}{3} \frac{\alpha_s}{r_{12}^3}
\right) + \frac{1}{m_1m_3} {\bf S}_1 \cdot ({\bf r}_{31}
\times {\bf p}_3) \left( \frac{2}{3} \frac{\alpha_s}{r_{31}^3}
\right) \right\} +
\nonumber\\
{} &+&
 \frac{1}{m_1m_2} \frac{2}{3} \alpha_s \left\{
\frac{1}{r_{12}^3} \left[ \frac{3}{r_{12}^2} ({\bf S}_1
\cdot {\bf r}_{12}) ({\bf S}_2 \cdot {\bf r}_{12}) -
{\bf S}_1 \cdot {\bf S}_2 \right] + \frac{8\pi}{3}
\delta^3 ({\bf r}_{12}) {\bf S}_1 \cdot {\bf S}_2
\right\}
\nonumber\\
{} &+& \, \mbox{cyclic permutations} \> ,
\end{eqnarray}
\begin{eqnarray}
V^{3q}_{{\rm vd}} &=&
 \sum_{j<l} \frac{1}{2m_jm_l} \left\{ \frac{2}{3}
\frac{\alpha_s}{r_{jl}} (\delta^{hk} + \hat{r}_{jl}^h \hat{r}_{jl}^k)
p_j^h p_l^k \right\}_{{\rm W}} -
\sum_{j=1}^{3} \frac{1}{6m_j^2} \{ \sigma \, r_j \,
 {\bf p}_{j{\rm T}_j}^2 \}_{{\rm W}} -
\nonumber\\
{} &-& \sum_{j=1}^{3} \frac{1}{6} \{ \sigma \, r_j \,
 \dot{{\bf z}}_{M{\rm T}_j}^2 \}_{{\rm W}} -
\sum_{j=1}^{3} \frac{1}{6m_j} \{ \sigma \, r_j \,
 {\bf p}_{j{\rm T}_j} \cdot \dot{{\bf z}}_{M{\rm T}_j} \}_{{\rm W}} \> .
\end{eqnarray}
Again ${\bf r}_{jl}={\bf z}_j - {\bf z}_l$ denotes the relative position
of the quark $j$  with respect to  the quark
$l$ ($j,l=1,2,3$) and ${\bf r}_j={\bf z}_j-{\bf z}_M$ the position
of the quark $j$ with respect to a common point $M$ such
that $\sum_{j=1}^3 r_j$ is minimum. As well known, if
no angle in the triangle made by the quarks exceeds $120^\circ$, the
three lines which connect the quarks with $M$ meet at this point with
equal angles of $120^\circ$ like a Mercedes star (I type
configuration, see Fig. 1A).
If one of the angles is $\geq 120^\circ$, then $M$
coincides with the respective vertex and the potential becomes
a two-body one (II type configuration, see Fig. 1B). Furthermore
$\nabla^2_{(j)}$ is the Laplacian with respect to the variable $z_j$
and
now $v^h_{{\rm T}_j}= (\delta^{hk}-\hat{r}_j^h \hat{r}_j^k)v^k$ where
$\hat{{\bf r}}_j = ({\bf r}_j/r_j)$.  The quantity $\dot{{\bf
z}}_{M}$
 in (1.8) is given by
\begin{equation}
 \dot{\bf z}_{M} = \left\{
\begin{array} {ll}
N^{-1} \sum_{j=1}^3 \left( {\bf p}_{j {\rm T}_j}/r_j  m_j \right)
\qquad \mbox{I type configuration}& \\
{\bf p}_{l}/m_l \qquad \qquad
\qquad \qquad \quad \mbox{ II type configuration}:
{\bf z}_M\equiv{\bf z}_l \> ,
\end{array} \right.
\end{equation}
$N$ being a matrix with elements $N^{hk}= \sum_{j=1}^3 \frac{1}{r_j}
(\delta^{hk} -\hat{r}^h_j \hat{r}^k_j)$.

Finally Eq. (1.7) properly refers to the  I configuration case.
 In general one should write
\begin{equation}
V_{\rm sd}^{\rm LR}= - \sum_{j=1}^3 {1 \over 2 m_j^2}
 {\bf S}_j \cdot {\bf \nabla}_{(j)} V_{\rm stat}^{\rm LR} \times
 {\bf p}_j \> .
\end{equation}
In comparing (1.10) with (1.7) one should keep in mind that the
partial derivatives in ${\bf z}_M$ of
$V_{\rm stat}^{\rm LR}$ vanish due to the definition of $M$.

We observe that
the short range part in Eqs. (1.6)-(1.8) is of a pure two body
type: it is identical to the electromagnetic
 potential among three equal charged
particles but for the colour group factor $2/3$ and
it is well known. Even the static
confining potential in Eq. (1.6) is known [7,1,3].
 The long range part of
Eq. (1.7) coincides with the expression obtained by Ford [8] starting
from the assumption of a purely scalar Salpeter potential of the form
\begin{equation}
 \sigma \, (r_1 + r_2 + r_3) \, {\beta}_1 {\beta}_2 {\beta}_3 \> ,
\end{equation}
but at our knowledge it has not been obtained consistently in a
Wilson loop context before. Eq. (1.8) is new. It should  be stressed that
(1.11) corresponds to the usual assumption of scalar confinement for the
quark-antiquark system. As well known from this assumption $V_{\rm
stat}^{q\overline{q}}$ and $V_{\rm sd}^{q\overline{q}}$ result identical
to (1.2) and (1.3), but $V_{\rm vd}^{q\overline{q}}$ turns out
different from (1.4).

The important point concerning Eqs. (1.1)-(1.8) and (1.10)
is that they follow from
rather reasonable assumptions on the behaviour  of two well
 known QCD objects, $W_{q \bar{q}}$ and $W_{3q}$, related to
 the appropriate (distorted) quark-antiquark and three-quark
 ``Wilson loops'' respectively. \par

For the $q\overline{q}$ case the basic object is
 \begin{equation}
W_{q\overline{q}} = \frac{1}{3} \left\langle {\rm Tr \, P}
 \exp \left( ig \oint_{\Gamma} dx^{\mu} \,
 A_{\mu}(x) \right) \right\rangle \> .
\end{equation}
Here the integration loop $\Gamma$ is assumed to be made by a world line
$\Gamma_1$ between an
initial position ${\bf y}_1$ at the time $ t_{\rm i} $ and a final one ${\bf
x}_1$ at the time
  $ t_{\rm f} $  for the quark ($t_{\rm i} < t_{\rm f}$), a similar world
line $\Gamma_2$ described
 in the reverse direction from ${\bf x}_2$ at the time $t_{\rm f}$
to ${\bf y}_2$ at the time $t_{\rm i}$ for the antiquark and two straight lines
at fixed times which  connect ${\bf x}_1$ to ${\bf x}_2$, ${\bf y}_2$
to ${\bf y}_1$ and close
the contour (Fig. 2). As usual $ A_{\mu} (x) =
\frac{1}{2} {\lambda}_ {a} A_{\mu} ^{a} (x) $,
  P prescribes the ordering
of the color matrices (from right to left) according to the direction fixed
on the loop and the angular brackets denote
 the functional integration
on the gauge fields.

The quantity $ i  \ln W_{q\overline{q}} $ is
written as the sum of a short range
 contribution (SR) and of a long range one (LR):
 $i\ln W_{q\overline{q}}=i\ln W_{q\overline{q}}^{{\rm
SR}} + i\ln W_{q\overline{q}}^{{\rm LR}}$. Then it is assumed that
the first term is given by the ordinary
perturbation theory, that is at the lowest
order
\begin{equation}
 i  \ln W^{{\rm SR}}_{q\overline{q}} =
 \frac{4}{3} g^2 \int_{\Gamma_1}dx^{\mu}_1
 \int_{\Gamma_2} dx^{\nu}_2 iD_{\mu \nu} (x_1-x_2)
\end{equation}
($ D_{\mu \nu} $ being the usual gluon propagator
 and ${\alpha_{s}}=g^2/4\pi$ the strong
interaction constant) and the second term by the so called ``area
law''
[9,1,4]
\begin{equation}
i \ln W^{{\rm LR}}_{q\overline{q}} = {\sigma} S_{\min} \> ,
\end{equation}
where $ S_{\min} $ denotes the minimal
 surface enclosed by the loop (${\sigma}$
is the string tension). Obviously Eq. (1.13) is justified by asymptotic
freedom,
Eq. (1.14) is suggested by  lattice theory, numerical simulation,
string models and other types of arguments.

Up to the $1/m^2$ order, the minimal surface can be identified with
the surface spanned by the straight line joining $(t, {\bf z}_1(t))$
to $(t, {\bf z}_2(t))$ with $t_{\rm i} \leq t \leq t_{\rm f}$; the generic
point of this surface is [4]
\begin{equation}
u^0_{\rm min}=t \quad \quad \quad
{\bf u}^{\min} = s {\bf z}_1(t) + (1-s) {\bf z}_2(t)
\end{equation}
with $0\leq s \leq 1$ and ${\bf z}_1(t)$ and ${\bf z}_2(t)$ being the
positions of the quark and the antiquark at the time $t$.

We further perform the so called instantaneous
approximation in (1.13), consisting in replacing
\begin{equation}
D_{\mu\nu}(x) \longrightarrow D_{\mu\nu}^{{\rm inst}}(x)= \delta(t)
\int_{-\infty}^{+\infty} d\tau \, D_{\mu\nu}(\tau, {\bf x})
\end{equation}
and use (1.15) and (1.16) at an early stage in the derivation
procedure. In this way we shall
obtain  Eqs. (1.1)-(1.4) in a much more direct way and without
the need of assuming a priori the existence of a potential as done
in [4]. So, once that Eqs. (1.13) and (1.14) have been written,
Eqs. (1.15) and (1.16) give the conditions
under which a description in terms of a potential actually holds.

Notice that, while (1.12), (1.14), and even (1.13) in the limit of large
$t_{\rm f}-t_{\rm i}$, are gauge invariant quantities, the error introduced by
(1.16) is strongly gauge dependent. The best choice of the gauge at the
lowest order in perturbation theory is the Coulomb gauge for which the
above error is minimum. To this choice Eq. (1.4) does refer.

For the three-quark case  the quantity analogous to (1.12) is
\begin{eqnarray}
  W_{3q} = \frac{1}{3!} \left\langle \varepsilon_{a_1 a_2 a_3}
 \varepsilon_{b_1 b_2 b_3}  \left[ {\rm  P} \exp \left( ig
 \int_{\overline{\Gamma}_1} dx^{\mu_1} A_{\mu_1}(x) \right)
\right]^{a_1 b_1} \right.
\nonumber\\
\left. \left[ {\rm P} \exp \left( ig \int_{\overline{\Gamma}_2}
 dx^{\mu_2} A_{\mu_2}(x) \right) \right]^{a_2 b_2}
  \left[  {\rm P} \exp \left( ig \int_{\overline{\Gamma}_3 } dx^{\mu_3}
A_{\mu_3}(x) \right) \right]^{a_3 b_3} \right\rangle \> .
\end{eqnarray}
Here $a_j,b_j$ are colour indices;
$\overline{\Gamma}_j$  denotes a curve made by: a
world line $\Gamma_j$ for the quark $j$
between the times $t_{\rm i}$ and $t_{\rm f}$
($t_{\rm i} < t_{\rm f}$), a straight  line
on the surface $t=t_{\rm i}$ merging from a point $I$ (whose
coordinate we denote by $y_M$) and connected to
the world line, another straight line
on the surface $t=t_{\rm f}$ connecting
the world line to a point $F$ with coordinate $x_M$ (Fig. 3).
The positions of the two points $I$
and $F$ are determined by the same rules which determine the point $M$
above.

The assumption corresponding to (1.13),(1.14) is then
\begin{equation}
 i \ln W_{3q} = \frac{2}{3} g^2 \sum _{i<j}
 \int _{\Gamma _i} dx^{\mu}
_i \int _{\Gamma _j} dx^{\nu}_j \,i D_{\mu \nu} (x_i - x_j)
+ \sigma S_{\min} \> ,
\end{equation}
where now $S_{\min}$ denotes the minimum among the surfaces made by three
sheets
having the curves $\overline{\Gamma}_1$,
 $\overline{\Gamma}_2$ and $\overline{\Gamma}_3$ as contours and
joining on a line $\Gamma_M$ connecting $I$ with $F$.

We shall see that Eqs. (1.5)-(1.8) follow
if we substitute (1.16) in (1.18) and
again replace $S_{\min}$ with the surface spanned by the straight
lines
\begin{equation}
u^0_{j \, \min}=t \quad \quad \quad
{\bf u}^{\min}_j = s {\bf z}_j(t) +(1-s) {\bf z}_M(t)
\end{equation}
with $j=1,2,3$, $s \in [0,1]$, ${\bf z}_M(t)$ being again the point
for which $\sum_{j=1}^3 |{\bf z}_j(t) - {\bf z}_ M(t)|$ is minimum.

The plan of the paper is the following one.
In Sec. II we shall report the simplified derivation of the
quark-antiquark potential as sketched above.
In Sec. III we shall report the derivation of the three-quark potential.
In Sec. IV we shall make some remarks and discuss the
connection with the flux tube model.

\section{Quark-antiquark potential}

As usual the starting point is the gauge invariant quark-antiquark
$(q_1,\bar{q}_2)$ Green function
(for the moment we assume the quark and the antiquark to have
 different flavours)
\begin{eqnarray}
G(x_1,x_2,y_1,y_2) &=&
\frac{1}{3}\langle0|{\rm T}\psi_2^c(x_2)U(x_2,x_1)\psi_1(x_1)
\overline{\psi}_1(y_1)U(y_1,y_2)  \overline{\psi}_2^c(y_2)
|0\rangle =
\nonumber\\
&=& \frac{1}{3} {\rm Tr} \langle U(x_2,x_1)
 S_1^{{\rm F}}(x_1,y_1|A) U(y_1,y_2) C^{-1}
S_2^{{\rm F}}(y_2,x_2|A) C \rangle \> .
\end{eqnarray}
Here $c$ denotes the charge-conjugate fields, $C$ is the charge-conjugation
matrix, $U$
the path-ordered gauge string
\begin{equation}
U(b,a)= {\rm P}  \exp  \left(ig\int_a^b dx^{\mu} \, A_{\mu}(x) \right)
\end{equation}
(the integration path being the straight line joining $a$ to $b$),
 $S_1^{{\rm F}}$ and $S_2^{{\rm F}}$ the quark propagators in an
external gauge field $A^{\mu}$; furthermore in principle the angular
brackets should be defined as
\begin{equation}
\langle f[A] \rangle = \frac{\int {\cal D}[A] M_f(A)
f[A] e^{iS[A]}}
{\int {\cal D}[A] M_f(A) e^{iS[A]}} \> ,
\end{equation}
$S[A]$ being the pure gauge field action and $M_f(A)$
the determinant resulting from the explicit integration on the
fermionic fields. In practice  assuming (1.14) corresponds
to take  $M_f(A)=1$ (quenched approximation).

Summarizing the first part of the procedure followed in Ref. [4]
(see such paper for details) first we assume $x_1^0=x_2^0=t_{\rm f}$,
$y_1^0=y_2^0=t_{\rm i}$ with $\tau=t_{\rm f}-t_{\rm i} >0$ and
note that $S_j^{{\rm F}}$ are $4\times4$  Dirac indices matrices type.
Then performing a Foldy-Wouthuysen transformation on $G$
we can replace $S_j^{{\rm F}}$ with a Pauli propagator $K_j$ (a
$2\times2$ matrix in the spin indices) and obtain a two-particle
Pauli-type
Green function $K$. We shall show that in the
described approximations this function satisfies a
Schr\"odinger-like equation  with the potential (1.1)-(1.4).

One finds (see [4]) that,
up to the $1/m^2$ order, $K_j$ satisfies the following equation
\begin{eqnarray}
& &i \frac{\partial}{\partial x^0} K_j(x,y|A)
 =  H_{{\rm FW}}
K_j(x,y|A) :=
\nonumber\\
& & =  \left[ m_j +\frac{1}{2m_j} ({\bf p}_j - g{\bf A})^2
 -  \frac{1}{8m_j^3} ({\bf p}_j -
g{\bf A})^4 - \frac{g}{m_j}
{\bf S}_j \cdot {\bf B} + gA^0 \right. -
\nonumber\\
& &{} - \left. \frac{g}{8m_j^2} (\partial_i E^i - ig [A^i,E^i])
 +  \frac{g}{4m_j^2} \varepsilon^{ihk} S_j^k
\{(p_j -gA)^i,E^h\} \right] K_j(x,y|A)
\end{eqnarray}
with the Cauchy condition
\begin{equation}
K_j(x,y|A) |_{x^0=y^0} = \delta^3({\bf x}-{\bf y})
\end{equation}
where $\varepsilon^{ihk}$ is the three-dimensional Ricci symbol and
the summation over repeated indices is understood.
By standard techniques the solution of Eq. (2.4), with the initial
condition (2.5), can be expressed as a path integral in phase space
\begin{equation}
K_j(x,y|A)= \int_{{\bf z}_j(y^0)={\bf y}}^{{\bf z}_j(x^0)={\bf x}}
{\cal D} [{\bf z}_j, {\bf p}_j] \, {\rm T}
\exp \left\{ i \int_{y^0}^{x^0}
dt \, [{\bf p}_j \cdot \dot{{\bf z}}_j - H_{{\rm FW}}
] \right\} \> ;
\end{equation}
here the time-ordering prescription T acts both on spin and gauge
matrices, the trajectory of the quark $j$ in configuration space is
denoted by ${\bf z}_j = {\bf z}_j(t)$, the trajectory in momentum
space by ${\bf p}_j = {\bf p}_j(t)$ and the spin by
${\bf S}_j$.
Then, by performing the translation
\begin{equation}
{\bf p} \longrightarrow {\bf p} +g{\bf A} \> ,
\end{equation}
we obtain an equation containing the expression $dt \, (gA^0 -
g \dot{{\bf z }} \cdot {\bf A} )
 \equiv g \,  dx^{\mu} A_{\mu}$, which is formally covariant.
\footnote{ More precisely, since the $A^h$
are matrices, the step $\int d^3{\bf p} \, f({\bf p}-g{\bf A})
= \int d^3{\bf p} \, f({\bf p})$ can be justified by expanding
$f({\bf p}-g{\bf A})$ in powers of $g$; apart from the zeroth order
term, all the other terms involve derivatives of $f({\bf p})$ and do
not contribute to the integral.}
It is also  useful to have an expression
for $K_j$ in which the tensor field $F^{\mu\nu}$ and its dual
$\hat{F}^{\mu\nu}$ appear. To this end we make the further translation
\begin{equation}
{\bf p} \longrightarrow {\bf p} -\frac{g}{m} ({\bf E} \times
{\bf S})
\end{equation}
and, apart from higher order terms, we obtain
\begin{eqnarray}
K_j(x,y|A)= \int_{{\bf z}_j(y^0)={\bf y}}^{{\bf z}_j(x^0)={\bf x}}
{\cal D} [{\bf z}_j,{\bf p}_j] \, {\rm T}
\exp \left\{ i \int_{y^0}^{x^0}
dt \, \left[ {\bf p}_j \cdot \dot{{\bf z}}_j - m_j -
\frac{{\bf p}_j^2}{2m_j} + \frac{{\bf p}_j^4}{8m_j^3} -
\right. \right.
\nonumber\\
{} - gA^0 + \frac{g}{m_j} {\bf S}_j \cdot {\bf B} +
\frac{g}{2m_j^2} {\bf S}_j \cdot ({\bf p}_j \times {\bf E}) -
\frac{g}{m_j} {\bf S}_j \cdot (\dot{{\bf z}}_j \times
{\bf E}) +
\nonumber\\
\left. \left. {} + g \dot{{\bf z}}_j \cdot {\bf A} +
\frac{g}{8m_j^2} (\partial_i E^i -ig[A^i,E^i])
\right] \right\} \> .
\end{eqnarray}

Thus we obtain the two-particle Pauli-type propagator $K$ in the form
of a path integral on the world lines of the two quarks
\begin{eqnarray}
K({\bf x}_1, {\bf x}_2, {\bf y}_1, {\bf y}_2;\tau)=
\int_{{\bf z}_1(t_{\rm i})={\bf y}_1}
    ^{{\bf z}_1(t_{\rm f})={\bf x}_1}
    {\cal D}  [{\bf z}_1, {\bf p}_1]
\int_{{\bf z}_2(t_{\rm i})={\bf y}_2}
    ^{{\bf z}_2(t_{\rm f})={\bf x}_2}
    {\cal D}   [{\bf z}_2, {\bf p}_2]
\nonumber\\
 \exp\left\{i\int_{t_{\rm i}}^{t_{\rm f}} dt\, \sum_{j=1}^2
\left[{\bf p}_j \cdot \dot{{\bf z}}_j -m_j-
\frac{{\bf p}^2_j}{2m_j}+\frac{{\bf p}^4_j}{8m_j^3}\right] \right\}
\nonumber\\
\left\langle \frac{1}{3}
{\rm Tr \, T_s \, P} \exp\left\{ig\oint_{\Gamma} dx^{\mu} \,
A_{\mu}(x) +\sum_{j=1}^2\frac{ig}{m_j} \int_{{\Gamma}_j} dx^{\mu}
\right. \right.
\nonumber\\
\left. \left. \left(S_j^l \hat{F}_{l{\mu}}(x) -\frac{1}{2m_j}
S_j^l\varepsilon^{lkr}p_j^k F_{{\mu}r}(x)-
\frac{1}{8m_j} D^{\nu}F_{{\nu}{\mu}}(x)
 \right) \right\} \right\rangle \> .
\end{eqnarray}
Here ${\rm T_s}$ is the time-ordering prescription for spin matrices, P
is the path-ordering prescription for gauge matrices along the loop
$\Gamma$ and as usual
\begin{equation}
F^{\mu\nu}= \partial^{\mu} A^{\nu} - \partial^{\nu} A^{\mu}
+ ig[A^{\mu},A^{\nu}] \> ,
\end{equation}
\begin{equation}
\hat{F}^{\mu\nu}= \frac{1}{2} \varepsilon^{\mu\nu\rho\sigma}
F_{\rho\sigma} \> ,
\end{equation}
\begin{equation}
D^{\nu}F_{\nu\mu}= \partial^{\nu}F_{\nu\mu}+
ig[A^{\nu},F_{\nu\mu}]
\end{equation}
and $\varepsilon^{\mu\nu\rho\sigma}$ is the four-dimensional Ricci
symbol.

Furthermore as in Eq. (1.12) $\Gamma_1$ denotes the path
going from $(t_{\rm i}, {\bf y}_1)$
to $(t_{\rm f},{\bf x}_1)$   along the quark
trajectory $(t,{\bf z}_1(t))$, $\Gamma_2$ the path going from
$(t_{\rm f}, {\bf x}_2)$ to $(t_{\rm i}, {\bf y}_2)$   along
the antiquark trajectory $(t, {\bf z}_2(t))$ and $\Gamma$ is the
path made by $\Gamma_1$ and $\Gamma_2$  closed by the
two straight lines joining $(t_{\rm i}, {\bf y}_2)$ with
$(t_{\rm i}, {\bf y}_1)$ and $(t_{\rm f}, {\bf x}_1)$
with $(t_{\rm f}, {\bf x}_2)$ (see Fig. 2). Finally Tr denotes
the trace on the gauge matrices. Note that the right-hand side of
(2.10) is manifestly gauge invariant.

What we have to show is that the angular bracket term in Eq.
(2.10) can be expressed as the exponential of an integral
function of the position,
momentum and spin alone taken at the same time $t$:
\begin{equation}
\left\langle \frac{1}{3}
{\rm Tr \, T_s \, P} \exp \ldots\right \rangle \simeq
{\rm T_s} \exp\left[ -i\int_{t_{\rm i}}^{t_{\rm f}} dt \,
 V^{q\overline{q}}({\bf z}_1,{\bf
z}_2,{\bf p}_1,{\bf p}_2,{\bf S}_1,{\bf S}_2) \right] \> ;
\end{equation}
indeed then we can conclude that
\begin{equation}
i\frac{\partial}{\partial t} K= \left[ \sum_{j=1}^2 \left(
m_j+ \frac{{\bf p}_j^2}{2m_j}-\frac{{\bf p}_j^4}{8m_j^3}
\right) +V^{q\overline{q}} \right] K \> ,
\end{equation}
$V^{q\overline{q}}$ playing the
role of a two-particle potential.
To this aim, expanding the logarithm of the left-hand side of (2.14)
up to $1/m^2$ order, we should have
\begin{eqnarray}
 i \ln W_{q\overline{q}}  +
i \sum_{j=1}^2 \frac{ig}{m_j} \int_{{\Gamma}_j}dx^{\mu}
\bigg( S_j^l \, \langle\langle \hat{F}_{l\mu}(x)
 \rangle \rangle -\frac{1}{2m_j} S_j^l
 \varepsilon^{lkr} p_j^k \, \langle\langle
F_{\mu r}(x) \rangle\rangle -
\nonumber\\
{} - \frac{1}{8m_j} \, \langle\langle
D^{\nu} F_{\nu\mu}(x) \rangle\rangle \bigg)
 - \frac{1}{2} \sum_{j,j^{\prime}} \frac{ig^2}{m_jm_{j^{\prime}}}
{\rm T_s} \int_{{\Gamma}_j} dx^{\mu} \, \int_{{\Gamma}_{j^{\prime}}}
dx^{\prime\sigma}
\, S_j^l \, S_{j^{\prime}}^k
\nonumber\\
\bigg( \, \langle\langle \hat{F}_{l \mu}(x)
 \hat{F}_{k \sigma}(x^{\prime})
\rangle\rangle - \, \langle\langle \hat{F}_{l \mu}(x) \rangle\rangle
\, \langle\langle \hat{F}_{k \sigma}(x^{\prime}) \rangle\rangle \bigg)
 \simeq  \left[\int_{t_{\rm i}}^{t_{\rm f}} dt
 \, V^{q\overline{q}} \right] \> ,
\end{eqnarray}
with the notation
\begin{equation}
\langle\langle f[A] \rangle\rangle = \frac{\frac{1}{3} \left\langle
{\rm Tr \, P}\left[ \exp ( ig\oint_{\Gamma}dx^{\mu}
 \, A_{\mu}(x) ) \right] f[A]
\right\rangle}
{\frac{1}{3} \left\langle {\rm Tr \, P}  \exp ( ig\oint_{\Gamma} dx^{\mu}
\, A_{\mu}(x) ) \right\rangle}
\end{equation}
and $W_{q\overline{q}}$ defined in Eq. (1.12).

At this point in Ref. [4]  we  assumed that
a quantity  $V^{q\overline{q}}$ satisfying (2.16) existed and derived
its form. Here we no longer make such an a priori
assumption  but start directly from (1.13) and (1.14).

\subsubsection{Contribution to the potential coming from
 $i \ln W_{q\overline{q}}$}

In the Coulomb gauge we have
\begin{eqnarray}
D_{00}(x)&=& \frac{i}{4\pi} \frac{1}{|{\bf x}|} \delta(t) \> ,
\\
D_{hk}(x)&=& (\delta_{hk} - {(\nabla^2)}^{-1} \partial_h \partial_k)
D_{\rm F}(x)=
\nonumber\\
&=& i \int \frac{d^4k}{(2\pi)^4} \frac{1}{k^2} \left(
\delta^{hk} - \frac{k^hk^k}{|{\bf k}|^2} \right)
e^{-ikx} \> ,
\\
D_{0k}(x) &=& D_{k0}(x) = 0
\end{eqnarray}
where $D_{\rm F}(x)= \int \frac{d^4k}{(2\pi^4)} \frac{i}{k^2}
e^{-ikx}$.

The pure temporal part $D_{00}(x)$ is already of instantaneous type,
for the pure spatial part we have the instantaneous limit
\begin{eqnarray}
D_{hk}^{{\rm inst}} (x) &=& \delta(t) \int_{-\infty}^{+\infty} d\tau \,
D_{hk}(\tau, {\bf x}) =
\nonumber\\
&=& - \delta(t) \frac{i}{8\pi |{\bf x}|} \left( \delta^{hk} +
\frac{x^hx^k}{|{\bf x}|^2} \right) \> .
\end{eqnarray}

Replacing (2.18),(2.20) and (2.21) in (1.13) we have
\begin{equation}
i \ln W^{{\rm SR}}_{q\overline{q}} = \int_{t_{\rm i}}^{t_{\rm f}} dt \,
\left\{ - \frac{4}{3} \frac{\alpha_s}{r} + \frac{1}{2}
\frac{4}{3} \frac{\alpha_s}{r} (\delta^{hk} + \hat{r}^h
\hat{r}^k) \dot{z}_1^h \dot{z}_2^k \right\} \> .
\end{equation}

Similarly, if we denote by $u^{\mu}=u^{\mu}(s,t)$ the equation of
any surface with contour $\Gamma$ ($s \in [0,1],\,
t \in [t_{\rm i},t_{\rm f}], \,
u^0(s,t)=t, \, {\bf u}(1,t)= {\bf z}_1(t),
\, {\bf u}(0,t)= {\bf z}_2(t) \,$), we
can write
\begin{eqnarray}
i \ln W^{{\rm LR}}_{q\overline{q}} =
\sigma S_{\min} & = & \sigma \min \int_{t_{\rm i}}^{t_{\rm f}}
dt \, \int_0^1  ds
 \, \left[-
\left( \frac{\partial u^{\mu}}{\partial t} \frac{\partial u_{\mu}}
{\partial t} \right) \left( \frac{\partial u^{\mu}}{\partial s}
\frac{\partial u_{\mu}}{\partial s} \right) + \left(
\frac{\partial u^{\mu}}{\partial t} \frac{\partial u_{\mu}}
{\partial s} \right)^2 \right]^{\frac{1}{2}} =
\nonumber\\
& = & \sigma \min \int_{t_{\rm i}}^{t_{\rm f}} dt \, \int_0^1
 ds \, \left|\frac{\partial
{\bf u}}{\partial s } \right| \left\{ 1-\left[ \left(
\frac{\partial {\bf u}}{\partial t} \right)_{\rm T} \right]^2 \right\}^
{\frac{1}{2}} \> ,
\end{eqnarray}
where the index ${\rm T}$ denotes the transverse part with respect to the
unit vector $\hat{s}$:
\begin{equation}
\hat{s} = \frac{\partial {\bf u}}{\partial s} \bigg/
\left| \frac{\partial {\bf u}}{\partial s } \right| \> .
\end{equation}

Then in approximation (1.15) we have
\begin{eqnarray}
\frac{\partial {\bf u}^{\min}}{\partial s} &=&
{\bf z}_1(t) - {\bf z}_2(t) \equiv {\bf r}(t) \> ,
\\
\frac{\partial {\bf u}^{\min}}{\partial t} &=&
s \dot{{\bf z}}_1(t) + (1-s) \dot{{\bf z}}_2(t)
\end{eqnarray}
and so
\begin{eqnarray}
i \ln W^{{\rm LR}}_{q\overline{q}} &=&
\int_{t_{\rm i}}^{t_{\rm f}} dt \, \sigma r \int_0^1 ds \, [1-(s
\dot{{\bf z}}_{1 \rm T} + (1-s)
 \dot{{\bf z}}_{2 \rm T} )^2]^{\frac{1}{2}} =
\nonumber\\
&=&  \int_{t_{\rm i}}^{t_{\rm f}} dt \, \sigma r \,
 \left[ 1-\frac{1}{6}
(\dot{{\bf z}}_{1 \rm T}^2
+ \dot{{\bf z}}_{2 \rm T}^2
+ \dot{{\bf z}}_{1 \rm T} \cdot \dot{{\bf z}}_{2 \rm T} ) + \ldots \, \,
\right]
\end{eqnarray}
where obviously now $\hat{s}=\hat{r}$ and we have expanded the square
root and performed the $s$ integration explicitly.

Notice now that, at the lowest order, $\dot{{\bf z}}_j$
can be replaced by ${\bf p}_j/m_j$ in (2.22) and (2.27). Then such
equations become of the correct form required by
(2.16) and so does the entire $i \ln W_{q\overline{q}}$.
In conclusion we have a first contribution
to $V^{q\overline{q}}$ (pure Wilson loop contribution) in the form
$V^{q\overline{q}}_{{\rm stat }} + V^{q\overline{q}}_{{\rm vd}}$
with $V_{\rm stat}^{q \bar{q}}$ and $V_{\rm vd}^{q \bar{q}}$ as
given in (1.2), (1.4). The ordering prescription in (1.4) shall
be discussed in Sec. IV.

\subsubsection{Spin related potential}

To obtain the remaining part of the potential we must
evaluate the  expectation values  of the form (2.17) occurring in (2.16).

Let us consider an arbitrary infinitesimal variation $ {\bf z}_1(t)
\longrightarrow {\bf z}_1(t) + \delta {\bf z}_1(t)$
vanishing at $t=t_{\rm f}$ and $t=t_{\rm i}$ and evaluate
 $\delta (i \ln W_{q\overline{q}})$.
 \footnote{The result is easily
achieved in the case of an abelian gauge theory: in fact one can freely
commute the fields $A_{\mu}$. In our case, on the contrary, the gauge
theory is non abelian: one can still commute fields referring to
different points because of the presence of the path-ordering operator
P; however, fields at the same point do not commute and this fact
must be taken into account. As a rule, one can make
calculations in the abelian case; by
rewriting the final result in a gauge invariant form, an expression
also valid in the non abelian case is obtained.}
{}From (1.12) we have
\begin{equation}
\delta W_{q\overline{q}} = - \frac{ig}{3}
 \left\langle {\rm Tr \, P} \int_{t_{\rm i}}^{t_{\rm f}}
\delta S^{\mu\nu}(z_1) \, F_{\mu\nu}(z_1) \,
\exp \left( ig \oint_{\Gamma} dx^{\mu} \, A_{\mu}(x) \right)
\right\rangle
\end{equation}
where $\delta S^{\mu\nu}(z_1) = \frac{1}{2} (dz_1^{\mu} \, \delta z_1^{\nu}
- dz_1^{\nu} \delta z_1^{\mu})$ is the element of the surface spanned
by $z_1(t)$.

Then
\begin{equation}
\delta (i \ln W_{q\overline{q}}) = i
 \frac{\delta  W_{q\overline{q}}}{W_{q\overline{q}}} =
g \int_{t_{\rm i}}^{t_{\rm f}} \delta S^{\mu\nu}(z_1)
 \, \langle \langle
F_{\mu\nu}(z_1) \rangle \rangle
\end{equation}
and we may write
\begin{equation}
g \, \langle \langle F_{\mu\nu}(z_1) \rangle \rangle =
\frac{ \delta (i \ln W_{q\overline{q}}) }
{ \delta S^{\mu\nu}(z_1)} \>
\end{equation}
(see App. A for a  definition of ${\delta \over \delta S^{\mu \nu}
(z_1)}$).
The computation is similar for the case of $z_2$ with a minus sign of
difference.

Similarly it can be seen that
\begin{equation}
g^2 \, \langle \langle F_{\mu\nu}(z_1) F_{\rho\sigma}(z_2) \rangle \rangle
= \frac{1}{W_{q\overline{q}}}
\frac{\delta^2 W_{q\overline{q}}}{ \delta S^{\mu\nu}(z_1)
\delta S^{\rho\sigma}(z_2) }
\end{equation}
and therefore
\begin{eqnarray}
g^2 \, \bigg( \langle \langle F_{\mu\nu}(z_1)
F_{\rho\sigma}(z_2) \rangle \rangle
- \langle \langle F_{\mu\nu}(z_1) \rangle \rangle \,
\langle \langle F_{\rho\sigma}(z_2)  \rangle \rangle \bigg) & = &
\nonumber\\
= \frac{ \delta^2 \ln W_{q\overline{q}}}
{\delta S^{\mu\nu}(z_1) \delta S^{\rho\sigma}(z_2)}
& = & - ig \frac{\delta}{\delta S^{\rho\sigma}(z_2)}
\, \langle \langle F_{\mu\nu}(z_1) \rangle \rangle \> .
\end{eqnarray}

{}From (1.13) we have then
\begin{equation}
g \, \langle\langle F_{\mu\nu}(z_1) \rangle\rangle^{{\rm SR}}
=  \frac{4}{3}
g^2 \int_{t_{\rm i}}^{t_{\rm f}} dt_2 \,
 i[ \partial_{\nu} D_{\mu\rho}(z_1-z_2)
- \partial_{\mu} D_{\nu\rho}(z_1-z_2) ] \dot{z}_2^{\rho}
\end{equation}
and
\begin{eqnarray}
g^2 \, \bigg( \langle\langle F_{\mu\nu}(z_1)
 F_{\rho\sigma}(z_2) \rangle\rangle
- \langle\langle F_{\mu\nu}(z_1) \rangle\rangle \,
\langle\langle F_{\rho\sigma}(z_2) \rangle\rangle \bigg)^{{\rm SR}} =
\nonumber\\
= \frac{4}{3} g^2 \{ \partial_{\rho} [ \partial_{\nu}
D_{\mu\sigma}(z_1-z_2) - \partial_{\mu} D_{\nu\sigma}(z_1-z_2)
] -
\nonumber\\
- \partial_{\sigma} [ \partial_{\nu} D_{\mu\rho}(z_1-z_2) -
\partial_{\mu} D_{\nu\rho}(z_1-z_2) ] \} \> .
\end{eqnarray}
Notice that  obviously in the terminology of App. A we have
$C(z_1, z_1')=0$ and then
\begin{equation}
\bigg( \langle \langle F_{\mu \nu} (z_1) F_{\rho\sigma} (z_{1}')\rangle
\rangle -\langle \langle  F_{\mu \nu }(z_1) \rangle \rangle \, \langle
\langle F_{\rho\sigma}(z_{1}') \rangle \rangle \bigg)^{SR}=0 \> .
\end{equation}

By using the Coulomb gauge and the instantaneous approximation we have
then
\begin{equation}
g \, \langle\langle F_{0k}(z_1) \rangle\rangle^{{\rm SR}} =
 \frac{4}{3} \alpha_s \frac{r^k}{r^3} \> ,
\end{equation}
\begin{equation}
g \, \langle\langle F_{hk}(z_1) \rangle\rangle^{{\rm SR}} =
 \frac{4}{3} \frac{\alpha_s}{m_2} \frac{1}{r^3}
( r^h p_2^k - r^k p_2^h ) \> ,
\end{equation}
\begin{eqnarray}
& & g^2 \, \bigg( \langle\langle F_{hk}(z_1) F_{lm}(z_2) \rangle\rangle -
\langle\langle F_{hk}(z_1) \rangle\rangle \,
\langle\langle F_{lm}(z_2) \rangle\rangle \bigg)^{{\rm SR}}  =
\nonumber\\
& &= - \frac{4}{3} \frac{ig^2}{8\pi} \delta (t_1-t_2) \left\{
\partial_l \partial_k \left[ \frac{1}{r} \left( \delta^{hm}
+ \hat{r}^h \hat{r}^m \right)\right] - \partial_l \partial_h \left[
\frac{1}{r} \left( \delta^{km} + \hat{r}^k
 \hat{r}^m \right) \right]
- \right.
\nonumber\\
& &\left. - \partial_m \partial_k \left[ \frac{1}{r} \left(
\delta^{hl} + \hat{r}^h \hat{r}^l \right)\right] + \partial_m
\partial_h \left[ \frac{1}{r} \left( \delta^{kl} + \hat{r}^k
\hat{r}^l \right)\right] \right\} \> .
\end{eqnarray}

In a similar way we obtain also
\begin{equation}
g \, \langle\langle F_{0k}(z_2) \rangle\rangle^{{\rm SR}} =
 \frac{4}{3} \alpha_s \frac{r^k}{r^3} \> ,
\end{equation}
\begin{equation}
g \, \langle\langle F_{hk}(z_2) \rangle\rangle^{{\rm SR}} =  \frac{4}{3}
\frac{\alpha_s}{m_1} \frac{1}{r^3} ( r^hp_1^k - r^kp_1^h ) \> .
\end{equation}

Let us now  consider the confinement part of $i \ln
W_{q\overline{q}}$. From Eq. (2.23), taking into account that $u^{\rm
min}_{\mu}$ satisfies the appropriate Euler equation, one has
(see App. B for details)
\begin{eqnarray}
g \, \langle\langle F_{\mu\nu}(z_1) \rangle\rangle^{{\rm LR}} =
\sigma \left[ \left( \frac{\partial u_{\mu}^{\min}}
{\partial s} \right)_1 \dot{z}_{1\nu} - \left(
\frac{\partial u_{\nu}^{\min}}{\partial s} \right)_1
\dot{z}_{1\mu} \right] \times
\nonumber\\
\times \left\{ - \dot{z}_1^2 \left( \frac{\partial u^{\min}}
{\partial s} \right)_1^2 + \left[ \dot{z}_1 \cdot \left(
\frac{\partial u^{\min}}{\partial s} \right)_1 \right]^2
\right\}^{-\frac{1}{2}}
\end{eqnarray}
where the subscript $1$ indicates that the derivative is calculated in
$s=1$. A similar formula is valid for $z_2$. Moreover as in the short
range case
\begin{equation}
\bigg( \langle\langle F_{\mu\nu}(z_1) F_{\rho\sigma}(z_1')
\rangle\rangle - \langle\langle F_{\mu\nu}(z_1) \rangle\rangle \,
\langle\langle F_{\rho\sigma}(z_1') \rangle\rangle \bigg)^{\rm LR}
=0 \> .
\end{equation}

By using the straight line approximation one  obtains
\begin{equation}
g \, \langle\langle F_{0k}(z_j) \rangle\rangle^{{\rm LR}} = \sigma
\frac{r^k}{r}+ O(v^2) \> ,
\end{equation}
\begin{equation}
g \, \langle\langle F_{hk}(z_j) \rangle\rangle^{{\rm LR}}
 = \frac{\sigma}{m_j}
\frac{1}{r} ( r^hp_j^k - r^kp_j^h )+O(v^3)
\end{equation}
with $j=1,2$ and
\begin{eqnarray}
g^2 \, \bigg( \langle\langle F_{hk}(z_1) F_{lm}(z_2)
\rangle\rangle - \langle\langle F_{hk}(z_1) \rangle\rangle
\, \langle\langle F_{lm}(z_2) \rangle\rangle
\bigg)^{{\rm LR}} &=& - ig {\delta \over \delta S^{lm}
(z_2)} \langle \langle F_{hk} (z_1) \rangle \rangle^{\rm LR}=
\nonumber\\
&=& O(v^2) \> .
\end{eqnarray}

Finally replacing (2.35)-(2.40) and (2.42)-(2.45) in (2.16)
we obtain the terms in Eq. (1.3) involving explicitly spin.

As concerns the Darwin-type terms one should evaluate $g \,
\langle\langle D^{\nu} F_{\nu\mu}(x) \rangle\rangle$ in Eq. (2.16).
By using the results of this Section one
can obtain by gauge invariance
\begin{eqnarray}
g \int_{\Gamma_1} dz_1^{\mu} \, \langle \langle D^{\nu} F_{\nu \mu}(z_1)
 \rangle \rangle &=& \int_{t_{\rm i}}^{t_{\rm f}} dt \, \dot{z}_1^{\mu}
\partial^{\nu} \langle \langle F_{\nu \mu} (z_1)\rangle \rangle =
\nonumber\\
= g \int_{t_{\rm i}}^{t_{\rm f}}
 dt \, \bigg( \partial^{\nu} \langle \langle F_{\nu 0}\rangle \rangle +
 \dot{z}_1^k \partial^{\nu} \langle \langle F_{\nu k} (z_1)\rangle
\rangle \bigg) &=&
\int_{t_{\rm i}}^{t_{\rm f}} dt \, \partial^h
\bigg(-{4\over 3}\alpha_s {r^h\over r^3} -
\sigma {r^h \over r} +\ldots \bigg) =
\nonumber\\
&=& \int_{t_{\rm i}}^{t_{\rm f}} dt \, \nabla^2
 \bigg(-{4\over 3} {\alpha_s \over r} +\sigma r \bigg) +\ldots
\end{eqnarray}
from which finally we have the Darwin terms as reported in Eq. (1.3).

\section{three-quark potential}

Let us define the three-quark color singlet state (again for three
 quarks of different flavours):
\begin{equation}
\frac{1}{\sqrt{3!}} \varepsilon_{b_1 b_2 b_3}
\overline{\psi}_{1 d_1}(y_1)
\overline{\psi}_{2 d_2}(y_2) \overline{\psi}_{3 d_3}(y_3)
U^{d_1 b_1}(y_1,y_M)
U^{d_2 b_2}(y_2,y_M) U^{d_3 b_3}(y_3,y_M)
| 0 \rangle
\end{equation}
where the path-ordered gauge strings $U$ are defined in Eq. (2.2),
$y_M$ is defined as explained after Eq. (1.17),
 $b_i$
and $d_i$ for $i=1,2,3$ are the color indices, $\varepsilon_{b_1 b_2 b_3}$
is the completely antisymmetric tensor in the color indices which
puts the system of three quarks in a color singlet state and
$\frac{1}{\sqrt{3!}}$ is the normalization factor.

The corresponding gauge invariant Green function  can be written
\begin{eqnarray}
& &G(x_1,x_2,x_3,y_1,y_2,y_3)
 = \frac{1}{3!} \varepsilon_{a_1 a_2 a_3}
\varepsilon_{b_1 b_2 b_3}
\nonumber\\
& &\langle 0| {\rm T} \,
U^{a_3 c_3}(x_M,x_3) U^{a_2 c_2}(x_M,x_2)
 U^{a_1 c_1}(x_M,x_1)
\psi_{3 c_3}(x_3)
\psi_{2 c_2}(x_2) \psi_{1 c_1}(x_1)
\nonumber\\
& &\overline{\psi}_{1 d_1}(y_1) \overline{\psi}_{2 d_2}(y_2)
\overline{\psi}_{3 d_3}(y_3)
U^{d_1 b_1}(y_1,y_M) U^{d_2 b_2}(y_2,y_M)
 U^{d_3 b_3}(y_3,y_M)
 |0 \rangle
\end{eqnarray}
and we assume
 $x_1^0=x_2^0=x_3^0= x_M^0= t_{\rm f}$,
$y_1^0=y_2^0=y_3^0= y_M^0=t_{\rm i}$;
$ \tau= t_{\rm f}- t_{\rm i} \, (\tau>0)$.

The integration over the Grassmann variables is again trivial
 and one can write
\begin{eqnarray}
G({\bf x}_1, {\bf x}_2, {\bf x}_3, {\bf y}_1, {\bf y}_2,
{\bf y}_3; \tau)
 = \frac{1}{3!} \varepsilon_{a_1 a_2 a_3}
\varepsilon_{b_1 b_2 b_3} \bigg\langle \bigg( U(x_M,x_1)
S_1^{\rm F}(x_1,y_1|A) U(y_1,y_M) \bigg)^{a_1 b_1}
\nonumber\\
\bigg( U(x_M,x_2) S_2^{\rm F}(x_2,y_2|A)
 U(y_2,y_M) \bigg)^{a_2 b_2}
\bigg( U(x_M,x_3) S_3^{\rm F}(x_3,y_3|A)
 U(y_3,y_M) \bigg)^{a_3 b_3}
\bigg\rangle \> ;
\end{eqnarray}
if some of the quarks are identical we have simply  to sum over all
 permutations of the corresponding  final variables.
 By performing even in this case the appropriate  Foldy-Wouthuysen
transformations and using (2.9)  we find in place  of (2.10)
\begin{eqnarray}
& &K({\bf x}_1, {\bf x}_2, {\bf x}_3, {\bf y}_1, {\bf y}_2,
{\bf y}_3; \tau | \, {\bf x}_M , {\bf y}_M ) =
\nonumber\\
& & =
\int_{{\bf z}_1(t_{\rm i})={\bf y}_1}^{{\bf z}_1(t_{\rm f})={\bf x}_1}
{\cal D} [{\bf z}_1,{\bf p}_1]
\int_{{\bf z}_2(t_{\rm i})={\bf y}_2}^{{\bf z}_2(t_{\rm f})={\bf x}_2}
{\cal D} [{\bf z}_2,{\bf p}_2]
\int_{{\bf z}_3(t_{\rm i})={\bf y}_3}^{{\bf z}_3(t_{\rm f})={\bf x}_3}
{\cal D} [{\bf z}_3,{\bf p}_3]
\nonumber\\
& &\exp \left\{ i \int_{t_{\rm i}}^{t_{\rm f}} dt \, \sum_{j=1}^{3}
\left[ {\bf p}_j \cdot \dot{{\bf z}}_j - m_j -
\frac{{\bf p}_j^2}{2m_j} + \frac{{\bf p}_j^4}{8m_j^3}
\right] \right\}
\nonumber\\
& &\left\langle \frac{1}{3!} \varepsilon \varepsilon
\prod_{j=1}^3 {\rm T_s \, P} \exp
\left\{ ig \int_{\overline{\Gamma}_j} dx^{\mu} \, A_{\mu}(x) +
\frac{ig}{m_j} \int_{{\Gamma}_j} dx^{\mu}
\right. \right.
\nonumber\\
& &\left. \left. \left( S_j^l \hat{F}_{l\mu}(x) -
\frac{1}{2m_j} S_j^l \varepsilon^{lkr} p_j^k F_{\mu r}(x) -
\frac{1}{8m_j} D^{\nu} F_{\nu\mu}(x)
\right) \right\} \right\rangle \> ,
\end{eqnarray}
$K$ being  now the three-quark Pauli-type Green function.

In (3.4) we have suppressed for convenience the color indices but
 have left trace of the tensors $\varepsilon_{a_1 a_2 a_3}
\varepsilon_{b_1 b_2 b_3}$ with the notation $\varepsilon
\varepsilon$.
As above ${\rm T_s}$ denotes the
 chronological ordering for the spin
matrices and  P is the path-ordering prescription acting  on the
gauge matrices; ${\Gamma}_j$  and $\overline{\Gamma}_j$
 are defined as in Eq. (1.17) and following.
 Notice that  the curve $\Gamma$ made by the union
 of $\overline{\Gamma}_1$, $\overline{\Gamma}_2$ and $\overline{\Gamma}_3$
is a closed three-branch loop  which
generalizes the Wilson loop of the two-body system.

{}From now on one can proceed  strictly as in Sec. II. We shall
 show  that from (1.18) using (1.16) and (1.19) one can write
\begin{equation}
\left\langle \frac{1}{3!} \varepsilon \varepsilon
\prod_{j=1}^3  {\rm T_s \, P} \exp
\ldots \right\rangle \simeq {\rm T_s} \exp
 \left[ -i \int_{t_{\rm i}}^{t_{\rm f}}
dt \, V^{3q}({\bf z}_1,{\bf z}_2,{\bf z}_3,{\bf p}_1,{\bf p}_2,{\bf
p}_3,{\bf S}_1,{\bf S}_2,{\bf S}_3) \right]
\end{equation}
and so the propagator $K$ obeys the three-particle
Schr\"{o}dinger-like equation
\begin{equation}
i \frac{\partial}{\partial t} K= \left[ \sum_{j=1}^{3} \left(
m_j + \frac{{\bf p}_j^2}{2m_j} - \frac{{\bf p}_j^4}{8m_j^3}
\right) + V^{3q} \right] K \> .
\end{equation}
Again expanding the logarithm of the left-hand side, Eq.(3.5)
turns to be equivalent to
\begin{eqnarray}
 i \ln W_{3q}  +
i \sum_{j=1}^3 \frac{ig}{m_j} \int_{{\Gamma}_j}dx^{\mu}
\bigg( S_j^l \, \langle\langle \hat{F}_{l\mu}(x)
 \rangle \rangle -\frac{1}{2m_j} S_j^l
 \varepsilon^{lkr} p_j^k \, \langle\langle
F_{\mu r}(x) \rangle\rangle -
\nonumber\\
{} - \frac{1}{8m_j} \, \langle\langle
D^{\nu} F_{\nu\mu}(x) \rangle\rangle \bigg)
 - \frac{1}{2} \sum_{j,j^{\prime}} \frac{ig^2}{m_jm_{j^{\prime}}}
{\rm T_s} \int_{{\Gamma}_j} dx^{\mu}
\, \int_{{\Gamma}_{j^{\prime}}}
dx^{\prime\sigma}
\, S_j^l \, S_{j^{\prime}}^k
\nonumber\\
\bigg( \, \langle\langle \hat{F}_{l \mu}(x)
 \hat{F}_{k \sigma}(x^{\prime})
\rangle\rangle - \, \langle\langle \hat{F}_{l \mu}(x) \rangle\rangle
\, \langle\langle \hat{F}_{k \sigma}(x^{\prime}) \rangle\rangle \bigg)
 \simeq  \left[\int_{t_{\rm i}}^{t_{\rm f}} dt
 \, V^{3q} \right]
\end{eqnarray}
with  $W_{3q}$ being defined as in Eq. (1.17) and now
\begin{equation}
\langle\langle f[A] \rangle\rangle= \frac{ \frac{1}{3!} \langle
\varepsilon \varepsilon \, \{ \prod_j \, {\rm P}
[ \exp (ig \int_{\overline{\Gamma}_j} dx^{\mu}
\, A_{\mu}(x) )] \} f[A] \rangle }{\frac{1}{3!} \langle \varepsilon
\varepsilon \, \{ \prod_j \, {\rm P} \exp
(ig \int_{\overline{\Gamma}_j} dx^{\mu} \, A_{\mu}(x)
) \} \rangle } \> .
\end{equation}

\subsubsection{Contribution to the potential coming from $i \ln W_{3q}$}

Having in mind (1.19)
let's evaluate $S_{\min}$ in the same manner as we did for the two-body
case. The quantity  $S_{\min}$ is the area made by three
  sheet surfaces as described in Sec. I.
Let us denote by $z_{M}(t)$ an arbitrary
 world line joining $y_M$ to $x_M$ and by $u_j^{\mu}=u_j^{\mu}(s,t)$ the
 equation of an arbitrary sheet interpolating between the trajectories
 $z_{j}^{\mu}=z_{j}^{\mu}(t)$ and $z_{M}^{\mu}=z_{M}^{\mu}(t)$;
obviously ${\bf u}_j(0,t)= {\bf z}_{M}(t)
, \, {\bf u}_j(1,t)={\bf z}_{j}(t)$ and $u_j^0(s,t)=t$.

Assuming  that the minimum  is taken  in the choice of $u^{\mu}_j(s,t)$ and
 of ${\bf z}_M(t)$ we can write  in analogy with (2.23)
\begin{equation}
S_{\min}= \min \sum_{j=1}^{3} \int_{t_{\rm i}}^{t_{\rm f}} dt \int_0^1 ds \,
\left| \frac{\partial {\bf u}_j}{\partial s} \right|
\left\{ 1- \left[ \left( \frac{\partial {\bf u}_j}{\partial t}
\right)_{{\rm T}_j} \right]^2 \right\}^{\frac{1}{2}}
\end{equation}
where now the index ${\rm T}_j$ stands
for transverse part of a vector with respect to
\begin{equation}
\hat{s}_j = \frac{\partial {\bf u}_j}{\partial s} \bigg/
\left| \frac{\partial {\bf u}_j}{\partial s} \right| \> .
\end{equation}

Then, performing the straight line approximation (1.19) we have
\begin{equation}
\frac{\partial {\bf u}_j^{\min} }{\partial s} = {\bf z}_j(t) -
{\bf z}_M(t) \equiv {\bf r}_j(t) \> ,
\end{equation}
\begin{equation}
\frac{\partial {\bf u}_j^{\min} }{\partial t} = s \dot{{\bf z}}_j
(t)+ (1-s) \dot{{\bf z}}_M(t)
\end{equation}
and expanding  in the velocities
\begin{equation}
S_{\min}= \int_{t_{\rm i}}^{t_{\rm f}} dt \, \sum_{j=1}^{3} r_j \left[ 1-
\frac{1}{6} (\dot{{\bf z}}_{j{\rm T}_j}^2
+ \dot{{\bf z}}_{M {\rm T}_j}^2
 + \dot{{\bf z}}_{j{\rm T}_j} \cdot \dot{{\bf z}}_{M {\rm T}_j})
+ \ldots \right]
\end{equation}
with $\hat{s}_j = \hat{r}_j$.

Taking into account  this result and introducing (2.18),(2.20) and
(2.21) in (1.18):
\begin{eqnarray}
i \ln W_{3q} = \int_{t_{\rm i}}^{t_{\rm f}} dt \, \left\{ \sum_{j<l}
\left[ - \frac{2}{3} \frac{\alpha_s}{r_{jl}} + \frac{1}{2} \frac{2}{3}
\frac{\alpha_s}{r_{jl}} (\delta^{hk} + \hat{r}_{jl}^h \hat{r}_{jl}^k)
\dot{z}_j^h \dot{z}_l^k \right] + \right.
\nonumber\\
\left. {}+ \sigma \sum_{j=1}^{3} r_j \left[ 1-\frac{1}{6}
 (\dot{{\bf z}}_{j{\rm T}_j}^2 + \dot{{\bf z}}^2_{M {\rm T}_j}
+ \dot{{\bf z}}_{j{\rm T}_j} \cdot
\dot{{\bf z}}_{M {\rm T}_j}) \right] \right\}
\end{eqnarray}
where again ${\bf r}_{jl} = {\bf r}_j -{\bf r}_l \equiv {\bf z}_j - {\bf
z}_l$. In the I configuration case
the quantity $\dot{\bf z}_M$ can be obtained deriving
 the equation $\sum_{j=1}^3 ({\bf r}_j / r_j)=0$. We get
 $\sum_{j=1}^3 {1\over r_j} (\delta^{hk} -\hat{r}_j^h
\hat{r}_j^k) \dot{z}_j^k =\sum_{j=1}^3 {1\over r_j} (\delta^{hk} -\hat{r}_j^h
\hat{r}_j^k) \dot{z}_M^k$. Obviously in the II configuration case we
have $\dot{\bf z}_M =\dot{\bf z}_l$ ($M \equiv {\rm quark} \, l$).

Finally replacing  $\dot{{\bf
z}}_j$ by ${\bf p}_j/m_j$  we  obtain Eq. (1.8)

\subsubsection{Spin related potential}

As in the $q \bar{q}$ case we can write
\begin{equation}
g \, \langle\langle F_{\mu\nu}(z_j) \rangle\rangle=
\frac{\delta (i \ln W_{3q} )}{\delta S^{\mu\nu}(z_j)} \> ,
\end{equation}
\begin{equation}
g^2 \, \bigg( \langle\langle F_{\mu\nu}(z_j) F_{\rho\sigma}(z_i)
\rangle\rangle - \langle\langle F_{\mu\nu}(z_j) \rangle\rangle \,
\langle\langle F_{\rho\sigma}(z_i) \rangle\rangle
\bigg) = ig \frac{\delta}{\delta
S^{\rho\sigma}(z_i) } \langle\langle
F_{\mu\nu}(z_j) \rangle\rangle
\end{equation}
with $j,i=1,2,3$
and adapt immediately the procedure used in the derivation of
Eqs. (2.35)-(2.40) and (2.42)-(2.45) to the variation of each single
quark world line separately.

{}From the short range part of (1.18) we have then
\begin{eqnarray}
g \, \langle\langle F_{\mu\nu}(z_j) \rangle\rangle^{{\rm SR}} = \frac{2}{3}
g^2 \left\{ \int_{t_{\rm i}}^{t_{\rm f}}
dt_i \, i[\partial_{\nu} D_{\mu\rho}
(z_j-z_i) - \partial_{\mu} D_{\nu\rho} (z_j-z_i) ]
\dot{z}_i^{\rho} + \right.
\nonumber\\
\left. + \int_{t_{\rm i}}^{t_{\rm f}} dt_n \,
i[ \partial_{\nu} D_{\mu\rho}
(z_j-z_n) - \partial_{\mu} D_{\nu\rho} (z_j-z_n) ]
\dot{z}_n^{\rho} \right\}
\end{eqnarray}
with $j,i,n=\, \mbox{cyclic permutation of} \, \, 1,2,3$ and
\begin{eqnarray}
g^2 \, \bigg( \langle\langle F_{\mu\nu}(z_j) F_{\rho\sigma}(z_i)
\rangle\rangle - \langle\langle F_{\mu\nu}(z_j) \rangle\rangle \,
\langle\langle F_{\rho\sigma}(z_i) \rangle\rangle \bigg)^{{\rm SR}} =
\nonumber\\
= \frac{2}{3} g^2 [\partial_{\sigma} \partial_{\nu}
D_{\mu\rho} (z_j-z_i) - \partial_{\sigma} \partial_{\mu}
D_{\nu\rho} (z_j-z_i) -
\nonumber\\
{} - \partial_{\rho} \partial_{\nu} D_{\mu\sigma}
(z_j-z_i) + \partial_{\rho} \partial_{\mu}
D_{\nu\sigma} (z_j-z_i) ]
\end{eqnarray}
for  $j \neq i$; furthermore
\begin{equation}
\bigg( \langle\langle F_{\mu\nu}(z_j) F_{\rho\sigma}(z_j^{\prime})
\rangle\rangle - \langle\langle F_{\mu\nu}(z_j) \rangle\rangle \,
\langle\langle F_{\rho\sigma}(z_j^{\prime}) \rangle\rangle
\bigg)^{{\rm SR}} = 0 \> .
\end{equation}

By using the Coulomb gauge and the instantaneous approximation we have
from Eqs. (3.17) and (3.18)
\begin{equation}
g \, \langle\langle F_{0k}(z_j) \rangle\rangle^{{\rm SR}} =  \frac{2}{3}
\alpha_s \left( \frac{r_{ji}^k}{r_{ji}^3} +
\frac{r_{jn}^k}{r_{jn}^3} \right) \> ,
\end{equation}
\begin{eqnarray}
g \, \langle\langle F_{hk}(z_j)
 \rangle\rangle^{{\rm SR}} & = & \frac{2}{3}
\frac{\alpha_s}{m_i} \frac{1}{r_{ji}^3} (r_{ji}^h p_i^k -
r_{ji}^k p_i^h) +
\nonumber\\
{} &+& \frac{2}{3} \frac{\alpha_s}{m_n} \frac{1}{r_{jn}^3}
(r_{jn}^h p_n^k - r_{jn}^k p_n^h)
\end{eqnarray}
($j,i,n= \, \mbox{cyclic permutation of} \, \, 1,2,3$) and
\begin{eqnarray}
& &g^2 \, \bigg( \langle\langle F_{hk}(z_j) F_{lm}(z_i) \rangle\rangle -
\langle\langle F_{hk}(z_j) \rangle\rangle \, \langle\langle
F_{lm}(z_i) \rangle\rangle \bigg)^{{\rm SR}} =
\nonumber\\
& & = -\frac{2}{3} \frac{ig^2}{8\pi} \delta(t_j-t_i) \left\{
\partial_l^{(i)} \partial_k^{(j)} \left[ \frac{1}{r_{ji}} \left(
\delta^{hm} + \hat{r}_{ji}^h \hat{r}_{ji}^m \right) \right] -
\partial_l^{(i)} \partial_h^{(j)} \left[ \frac{1}{r_{ji}} \left(
\delta^{km} + \hat{r}_{ji}^k \hat{r}_{ji}^m \right) \right]
- \right.
\nonumber\\
& & \left. {}- \partial_m^{(i)} \partial_k^{(j)} \left[
\frac{1}{r_{ji}} \left( \delta^{hl} + \hat{r}_{ji}^h
\hat{r}_{ji}^l \right) \right] + \partial_m^{(i)} \partial_h^{(j)}
\left[ \frac{1}{r_{ji}} \left( \delta^{kl} + \hat{r}_{ji}^k
\hat{r}_{ji}^l \right) \right] \right\}
\end{eqnarray}
where  $j \neq i$.

Let us now consider the long range part of Eq. (1.18).
Notice that, due to its definition, an infinitesimal variation of the
intermediate point world line ${\bf z}_M = {\bf z}_M(t)$ leaves the
quantity $i \ln W_{3q}$ unchanged. Then, in the evaluation of the
functional derivatives (3.15) and (3.16), we can treat such world line
as fixed.
{}From Eq. (3.9),
taking into account that $u_{j\mu}^{\min}$ satisfies the appropriate
Euler equation, one has
\begin{eqnarray}
g \, \langle\langle F_{\mu\nu}(z_j) \rangle\rangle^{{\rm LR}} =
\sigma \left[ \left( \frac{\partial u_{j\mu}^{\min}}
{\partial s_j} \right)_1 \dot{z}_{j\nu} - \left( \frac{\partial
u_{j\nu}^{\min}}{\partial s_j} \right)_1 \dot{z}_{j\mu}
\right] \times
\nonumber\\
{} \times \left\{ - \dot{z}_j^2 \left( \frac{\partial u_j^{\min}}
{\partial s_j} \right)_1^2 + \left[ \dot{z}_j \cdot \left(
\frac{\partial u_j^{\min}}{\partial s_j} \right)_1 \right]^2
\right\}^{- \frac{1}{2}}
\end{eqnarray}
 where the subscript $1$ indicates that the
derivative is calculated in $s=1$.

Moreover
\begin{equation}
\bigg( \langle\langle F_{\mu\nu}(z_j) F_{\rho\sigma}(z_j^{\prime})
\rangle\rangle - \langle\langle F_{\mu\nu}(z_j) \rangle\rangle \,
\langle\langle F_{\rho\sigma}(z_j^{\prime}) \rangle\rangle
\bigg)^{{\rm LR}} = 0 \> .
\end{equation}

By using the straight line approximation one obtains
\begin{equation}
g \, \langle\langle F_{0k}(z_j) \rangle\rangle^{{\rm LR}} =
\sigma \frac{r_j^k}{r_j} \> ,
\end{equation}
\begin{equation}
g \, \langle\langle F_{hk}(z_j) \rangle\rangle^{{\rm LR}} =
\frac{\sigma}{m_j} \frac{1}{r_j} (r_j^h p_j^k - r_j^k p_j^h)
\end{equation}
with $j=1,2,3$ and
\begin{eqnarray}
g^2 \, \bigg( \langle\langle F_{hk}(z_j) F_{lm}(z_i)
\rangle\rangle - \langle\langle F_{hk}(z_j) \rangle\rangle
\, \langle\langle F_{lm}(z_i) \rangle\rangle
\bigg)^{{\rm LR}} &=& ig \frac{\delta}{\delta
S^{lm}(z_i)} \langle\langle F_{hk}(z_j) \rangle\rangle^{{\rm LR}} =
\nonumber\\
&=& O(v^2)
\end{eqnarray}
with $j\neq i$.

As concerns the Darwin-type terms we must evaluate $g \, \langle\langle
D^{(j)\nu} F_{\nu\mu}(z_j) \rangle\rangle$ for $j=1,2,3$ with
$D^{(j)\nu} = \partial^{(j)\nu} + ig A^{\nu}$. This can be done
by using the same line
of derivation as  for the two-body case.

At this point we may derive the spin dependent potential:
after some calculations using Eqs. (3.19)-(3.22) and (3.24)-(3.27) in
Eq. (3.7) we get Eq. (1.7).

\section{Additional considerations}

To complete the work some additional considerations and remarks
 are necessary.

   1)   In Sec. III we have assumed different flavours for the three quarks.
If two or three quarks are identical we
must identify the corresponding  operators in Eq. (3.1) and add the
appropriate normalization factor ${1\over \sqrt{2!}}$ or ${1\over
 \sqrt{3!}}$. Correspondingly we
 have  to replace the right-hand
side in (3.3) with the corresponding sum over the permutations of the final
identical particles divided by the  combinatorial factor (2! or
3!). For what concerns the potential this means
 only that we have  to equate
the masses of such particles in (1.7) and (1.8). Similarly if in Sec. II the
quark and the antiquark have the same flavour, a new term must be added to
the last member of (2.1), which is obtained by interchanging the roles of
$y_1$ and $x_2$. Such ``annihilation'' term is not properly of potential type
but can be treated perturbatively [10].

   2)   Concerning the ordering problem in (1.4) and (1.8) we recall that
there are two independent possible prescriptions for a  quantity
 quadratic in the momenta: \par \noindent
the Weyl prescription
\begin{equation}
 \{ f({\bf r}) p^h p^k \}_{\rm W}
= {1\over 4} \{ \{ f({\bf r}), p^h  \} , p^k \}
\end{equation}
and the symmetric one
\begin{equation}
 \{ f({ \bf r}) p^h p^k \}_{\rm S} = {1\over 2}  \{ f({\bf r}),p^h p^k \} \>.
\end{equation}
As well known, in the path integral formalism the ordering
 prescription corresponds
to the specific discretization rules used in the definition of it.

In Ref. [5] we adopted a somewhat ad hoc rule
 corresponding to
the ordering $ \{ \, \, \}_{\rm ord} = {2 \over 3} \{ \, \, \}_{\rm W}
 + {1 \over 3} \{ \, \, \}_{\rm S} $.
Such rule was motivated by the fact that it enables a  by part
 integration at the discrete level which was necessary  to
 eliminate  a dependence of the potential  from the acceleration.

Notice however that the limit procedure  used for the definition
 of the path integrals in (2.9) or (2.10) is not at our
 choice but it is a consequence  of the corresponding procedure
used in the definition of the field functional  integration  in
 (2.1). To see what is  the correct prescription, let us assume
 a definite lattice with spacing $\bf{\varepsilon}$ in the time
 direction and a   spacing $ a$ in the space directions.
 Let us consider  the corresponding discrete counterpart of (2.1),
 written according to the usual rules for gauge theories [11] and
 perform the integration  of the fermionic fields at this discrete
 level. Then in place  of (2.10) we arrive  at an equation
  in which not only the time integral  is replaced  by a sum over the
appropriate  discrete  times $t_s$, but for every $t_s$ even the
 integrals on ${\bf z}_1$ and ${\bf z}_2$ are replaced  by the sum over
all the sites of the lattice corresponding  to that time coordinate.
 Finally Eq. (1.12) has to be interpreted as
\begin{equation}
W_{q \bar{q}} \simeq \int \prod_{\{n',n\}} {\rm d} U_{n'n}
 \, \,
e^{i S[U]} \, {\rm Tr} \, {\rm P} \prod_{\{r,r'\} \in \Gamma  } U_{r'r}
\end{equation}
where $U_{n' n}$ denotes the element of the color group  associated
 to the link between the contiguous sites $n$ and $n'$ and the product
is extended  to all such  links or to all links laying on the curve
$\Gamma$.
Since in turn $U_{n' n}$ can be  interpreted as
$\exp \left[ig(x_{n'}-x_{n})^{\mu} A_{\mu}({x_{n'} + x_{n} \over 2})
\right] $, we see that, after having explicitly  integrated
 over $U$ (and so used (1.13), (1.14)) and performed the limit
$a \to 0$, we are left with the discrete  form of an ordinary
path integral with
\begin{equation}
 X^{hk}({\bf r}) p_i^h p_j^k \longrightarrow X^{hk}({{\bf r}_s +
 {\bf r}_{s-1} \over
2})
          p_{is}^h p_{js}^k  \> .
\end{equation}
Eq. (4.4) does correspond to the Weyl ordering,
as indicated in (1.4) [12].
 Notice, however, that a different
ordering in (1.4) would bring simply
 to additional Darwin-like terms which in practice can be nearly
completely compensated by a readjustment of the potential parameters.
Similar arguments apply to Eq. (1.8) in the $3 q$ case.

3) To clarify the connection between the $ q \overline{q} $ potential and the
{\it relativistic flux tube model} [2] it is convenient to neglect
the  spin dependent terms in (2.10) and replace the $1/m^2$
 expansion
by its exact relativistic expression. We have
\begin{eqnarray}
& &K({\bf x}_1, {\bf x}_2;  {\bf y}_1, {\bf y}_2; t_{\rm f} - t_{\rm i}) =
\nonumber \\
& &\int {\cal D} [{\bf z}_1, {\bf p}_2]
\int {\cal D} [{\bf z}_2, {\bf p}_2]
\exp \left\{i \left[ \int_{t_{\rm i}}^{t_{\rm f}} dt \sum_{j=1}^2 ({\bf p}_j
\cdot {\dot{\bf z}_j} - \sqrt{m_j^2 + {\bf p}_j^2})
\right] + \ln W_{q \bar{q}} \right\} \> .
\end{eqnarray}
Further by taking advantage of (2.22) and of the first step in (2.27),
after expanding again the exponential in (4.5)
 around the values $ {\bf p}_j = m {\dot {\bf z}}_j /
 \sqrt {1 - {\dot {\bf z}}_j^2 } $ and performing the integration in the
gaussian approximation (semiclassical approximation), we can write
(see however in this connection ref[12])
\begin{eqnarray}
 K( {\bf x}_1 , {\bf x}_2  ; {\bf y}_1 ,& & {\bf y}_2 ,
 t_{\rm f} - t_{\rm i} ) =\nonumber \\
& &\int {\cal D} [ {\bf z}_1 ] \Delta [ {\bf z}_1 ] \int {\cal D} [ {\bf z}_2 ]
\Delta [ {\bf z}_2 ] \exp \left\{i \int_{t_{\rm i} }^{t_{\rm f}}
dt L( {\bf z}_1 ,
{\bf z}_2 , \dot {\bf z}_1 , \dot {\bf z}_2 ) \right\}
\end{eqnarray}
with
\begin{eqnarray}
 L \, & & = - \sum_{j=1}^2 m_j \sqrt{1- \dot {\bf z}_j^2 } +
 {4 \over 3} { {\alpha}_s  \over r} \left[ 1 - \frac{1}{2}
( \delta^{hk} + \hat{r}^h \hat{r}^k ) \dot{z}_1^h \dot{z}_2^k \right]
+\nonumber \\
 & & {} - \sigma
r \int_0^1 ds [1 - ( s \dot{\bf z}_{1{\rm T}} + (1-s) \dot{{\bf
z}}_{2{\rm T}} )^2 ]^{1/2}.
\end{eqnarray}
In (4.6) $ {\cal D} [ {\bf z} ] $ denotes the usual nonrelativistic
configurational measure and $ \Delta [ {\bf z} ] $
is a determinantal factor which
has to be be considered part of the relativistic measure. Formally we can write
\begin{equation}
{\cal D} [ {\bf z} ] = \left( {m \over 2 \pi i \epsilon} \right)^{3/2}
\prod_t \, \left[ \left(
{m \over 2 \pi i \epsilon} \right)^{3/2} d^3 {\bf z} (t) \right]
\end{equation}
($ \epsilon =$ time lattice spacing) and
\begin{equation}
\Delta [ {\bf z} ]= \left\{ \prod_t \det
 \left[{1 \over m} (1 - \dot{\bf z}^2)^{1/2}
( \delta^{hk}-{\dot{z}}^h {\dot{z}}^k ) \right]
\right\}^{-\frac{1}{2}} \> .
\end{equation}
What we want to stress is that
Eq. (4.7) is identical to the center of mass lagrangian
for the relativistic
flux tube model. This is consistent with what already observed at the
$ 1/m^2 $ order in Ref. [2].

4) In phenomenological analysis the following long
range static  potential of two-body type has been often adopted
[7] for the $3q$ case:
\begin{equation}
V_{{\rm stat}}^{{\rm LR}} = \frac{1}{2}
\sigma \, (r_{12} + r_{23} + r_{31})
\end{equation}
with  a  corresponding spin dependent  potential again of the form
(1.10)
\begin{equation}
V_{{\rm sd}}^{{\rm LR}} =
- \sum_{j=1}^3 \frac{1}{2m_j^2} {\bf S}_j \cdot
{\bf {\nabla}}_{(j)} V^{{\rm LR}}_{{\rm stat}} \times {\bf p}_j \> .
\end{equation}
The factor $1/2$ in Eq. (4.10) is motivated by the fact that when two quarks
collapse they become equivalent from the colour point of view to an
antiquark and $V_{\rm stat}^{3q}$, $V_{\rm sd}^{3q}$  must reduce to
$V^{q \bar{q}}_{\rm stat}$ and $V_{\rm sd}^{q \bar{q}}$.
 It has been shown
that (4.10) and (4.11) produce a spectrum very close to that obtained from
(1.6) and (1.10). From a numerical point of view
 the use of (1.6) amounts
  to replace the factor
$1/2$ in (4.10) by a factor of the order of  0.54-0.55
[7,8]. Notice however that from a fundamental point of view (4.10) has no
clear  basis.

5) As we already mentioned  in lattice gauge theory
 the area law is obtained  under
the approximation in which the quantity $ M_f(A) $ is replaced by 1, the so
called quenched approximation. Our entire treatment has to be understood
in this perspective. Then the effect of virtual quark-antiquark
creation should be introduced as a correction at a  later stage. Various
attempts have been done in this direction: see references [13,14].

6) In Eqs. (1.6)-(1.8)
$V_{{\rm sd}}^{3q}$ has been written in the coordinates ${\bf r}_1,
{\bf r}_2, {\bf r}_3$ and ${\bf r}_{12}, {\bf r}_{23},
{\bf r}_{31}$ which provides a symmetric form of the potential.
For numerical computations it may be useful to express such variables in
terms of a system of independent coordinates like the Jacobi
 coordinates (see [3] and [8]).

\section{Conclusions}

   In conclusion we have strongly simplified the derivation of the
quark-antiquark potential as given in [4,5]. We have shown that, once the
assumptions (1.13) and (1.14) are done, the existence of a potential follows
when one performs the instantaneous approximation (1.16) and the
straight line approximation (1.15). We have also  corrected the
 ordering prescription.

   Due to the above simplifications, the method has been extended without
difficulty
to the three-quark case, where the relevant assumptions and
approximations are
(1.18), (1.16) and (1.19). As a result a $3q$ spin dependent potential has been
consistently obtained in the Wilson loop context.
This   coincides with the
one already
proposed by Ford under an assumption of scalar confinement. It has also been
evaluated the $ O(1/m^2) $ $3q$
 velocity dependent potential which is new at our
knowledge.

   Notice that in both the $q \overline q $ and $ 3q $ cases the spin
independent relativistic corrections obtained by us differ from those
resulting
from the mentioned  assumption of scalar confinement and seems to
agree better with the data [14,10] (for the difficulties of the scalar
confinement hypothesis see also [15,16]).\par
Finally we have seen  that if in Eq. (2.10) we replace the kinetic
 terms by the exact relativistic expressions and neglect the spin
 dependent terms, if further we use
 the first step in Eq. (2.27) (without any velocity
 expansion) and perform the momentum integration in the
 semiclassical approximation, we obtain the lagrangian for the
 relativistic flux tube model [2].

\appendix
\section{}

Concerning the definition of the symbol $\frac{\delta}{\delta
S^{\mu\nu}(z)}$, let us consider a functional of the world line
$\gamma$ of the form
\begin{equation}
F[\gamma]=A+ \int_{\gamma} dz^{\mu} \, B_{\mu}(z) + \frac{1}{2}
\int_{\gamma} dz^{\mu} \int_{\gamma} dz^{\prime\rho} \,
C_{\mu\rho}(z,z^{\prime}) + \ldots
\end{equation}
with $C_{\mu\rho}(z,z^{\prime})=C_{\rho\mu}(z,z^{\prime})$ etc. Let us
denote by $z=z(\lambda)$ the parametric equation of $\gamma$ and
consider an infinitesimal variation of such curve $z(\lambda)
\longrightarrow z(\lambda)+ \delta z(\lambda)$. We have
\begin{eqnarray}
\delta F&=& \int_{\gamma} \left( \delta dz^{\mu} B_{\mu}(z) +
dz^{\mu} \frac{\partial B_{\mu}(z)}{\partial z^{\nu}} \delta z^{\nu}
\right) +
\nonumber\\
&+& \int_{\gamma} dz^{\prime\rho} \int_{\gamma} \left( \delta dz^{\mu}
C_{\mu\rho}(z,z^{\prime}) + dz^{\mu} \frac{\partial
C_{\mu\rho}(z,z^{\prime})}{\partial z^{\nu}} \delta z^{\nu}
\right) + \ldots \, =
\nonumber\\
&=& \frac{1}{2} \int_{\gamma} (dz^{\mu} \delta z^{\nu} - dz^{\nu} \delta
z^{\mu}) \left[ \left( \frac{\partial B_{\mu}(z)}{\partial z^{\nu}} -
\frac{\partial B_{\nu}(z)}{\partial z^{\mu}} \right) + \right.
\nonumber\\
&+& \left. \int_{\gamma}
dz^{\prime\rho} \left( \frac{\partial C_{\mu\rho}(z,z^{\prime})}{\partial
z^{\nu}} - \frac{\partial C_{\nu\rho}(z,z^{\prime})}{\partial
z^{\mu} } \right) + \ldots \, \right],
\end{eqnarray}
where a partial integration has been performed at the second step.
Assuming $\delta z(\lambda)$ different from zero only in a small
neighbourhood of a specific value $\overline{\lambda}$ of $\lambda$ we
write
\begin{equation}
\frac{\delta F}{\delta S^{\mu\nu}(\overline{z})} = \frac{\partial
B_{\mu}(\overline{z})}{\partial z^{\nu}} - \frac{\partial
B_{\nu}(\overline{z})}{\partial z^{\mu}} + \int_{\gamma}
dz^{\prime\rho} \left( \frac{\partial
C_{\mu\rho}(\overline{z},z^{\prime})}{\partial z^{\nu}} -
\frac{\partial C_{\nu\rho}(\overline{z},z^{\prime})}{\partial
z^{\mu}} \right) + \ldots
\end{equation}
with $\overline{z}= z(\overline{\lambda})$. Furthermore if we consider
a second variation $\delta z(\lambda)$ different from zero only in a
small neighbourhood of a second value $\overline{\lambda}'\neq
\overline{\lambda}$ we have
\begin{eqnarray}
\delta \frac{\delta F}{\delta S^{\mu\nu}(\overline{z})} = \frac{1}{2}
\int_{\gamma}(dz'^{\rho} \delta z'^{\sigma} - dz'^{\sigma} \delta
z'^{\rho}) \left[ \frac{\partial}{\partial z'^{\sigma}} \left(
\frac{\partial C_{\mu\rho}(\overline{z},z')}{\partial z^{\nu}} -
\frac{\partial C_{\nu\rho}(\overline{z},z')}{\partial z^{\mu}}
\right) - \right.
\nonumber\\
\left. - \frac{\partial}{\partial z'^{\rho}} \left( \frac{\partial
C_{\mu\sigma}(\overline{z},z')}{\partial z^{\nu}} - \frac{\partial
C_{\nu\sigma}(\overline{z},z')}{\partial z^{\mu}} \right) + \ldots \,
\right]
\end{eqnarray}
and
\begin{eqnarray}
\frac{\delta^2 F}{\delta S^{\rho\sigma}(\overline{z}') \delta
S^{\mu\nu}(\overline{z})} &=& \frac{\partial^2 C_{\mu\rho}(\overline{z},
\overline{z}')}{\partial z'^{\sigma} \partial z^{\nu}} -
\frac{\partial^2 C_{\nu\rho}(\overline{z},\overline{z}')}{\partial
z'^{\sigma} \partial z^{\mu}} -
\nonumber\\
{}&-& \frac{\partial^2 C_{\mu\sigma}(\overline{z},\overline{z}')}{\partial
z'^{\rho} \partial z^{\nu}} + \frac{\partial^2 C_{\nu\sigma}
(\overline{z},\overline{z}')}{\partial z'^{\rho} \partial z^{\mu}}
+ \ldots
\end{eqnarray}
Notice that the assumption $\bar{\lambda}' \ne \bar{\lambda}$  is
essential to make the definitions non ambiguous. The case in which
a $\delta(z-z')$ term occurres  in $C(z,z')$  must be treated as a
limit one. In practice  this amounts to say that (A3) and (A5)
hold true even in this case.

\section{}

In applying (A3) and (A5)  to $i \ln W_{q \bar{q}}$ [17] and $i \ln W_{3q}$
and specifically to Eqs. (2.23) and (3.9) it is
 convenient to think of $t$ as an
arbitrary parameter on the same foot of $\lambda$.
 It is obviously  understood that  at the end   $t$ is
identified with the ordinary time by setting $z_j^0(t)=t$ and
$u^0(s,t)=t$ (this in the $q\bar{q}$ case).
In particular  rewriting Eq. (2.23) as
\begin{equation}
i \ln W_{q\overline{q}}^{\rm LR} = \sigma S_{\min} = \sigma
\int_{t_{\rm i}}^{t_{\rm f}} dt \int_0^1 ds \, {\cal S}(u^{\min})
\end{equation}
with
\begin{equation}
{\cal S}(u)=
\left[-
\left( \frac{\partial u^{\mu}}{\partial t} \frac{\partial u_{\mu}}
{\partial t} \right) \left( \frac{\partial u^{\mu}}{\partial s}
\frac{\partial u_{\mu}}{\partial s} \right) + \left(
\frac{\partial u^{\mu}}{\partial t} \frac{\partial u_{\mu}}
{\partial s} \right)^2 \right]^{\frac{1}{2}} \>
\end{equation}
one can
notice that the equation of the minimal surface $u^{\mu}=
u^{\mu}_{\min}(s,t)$ is the solution of the Euler equations
\begin{equation}
\frac{\partial}{\partial s} \frac{\partial {\cal S}}{\partial
\left(\frac{\partial u^{\mu}}{\partial s} \right) } +
\frac{\partial}{\partial t} \frac{\partial {\cal S}}{\partial
\left(\frac{\partial u^{\mu}}{\partial t} \right) } =0
\end{equation}
satisfying the contour conditions $u^{\mu}(1,t)=z_1^{\mu}(t),
\, u^{\mu}(0,t)=z_2^{\mu}(t)$. Then considering an infinitesimal
variation of the world line of the quark 1, $z_1^{\mu}(t)
\longrightarrow z_1^{\mu}(t) + \delta z_1^{\mu}(t)$, even
$u^{\mu}_{\min}(s,t)$ must change, $u^{\mu}_{\min}(s,t)
\longrightarrow u^{\mu}_{\min}(s,t) + \delta u^{\mu}(s,t)$ and one has
\begin{eqnarray}
\delta (i \ln W_{q\overline{q}}^{\rm LR} ) &=& \sigma
\int_{t_{\rm i}}^{t_{\rm f}} dt \int_0^1 ds \, \left[  \frac{\partial
{\cal S}}{\partial \left(\frac{\partial u^{\mu}}{\partial s} \right) }
\frac{\partial}{\partial s} \delta u^{\mu} + \frac{\partial
{\cal S}}{\partial \left(\frac{\partial u^{\mu}}{\partial t} \right) }
\frac{\partial}{\partial t} \delta u^{\mu} \right]_{u=u_{\min}} =
\nonumber\\
&=& \sigma \int_{t_{\rm i}}^{t_{\rm f}} dt \, \left[ \frac{\partial
{\cal S}}{\partial \left(\frac{\partial u^{\mu}}{\partial s} \right) }
\delta u^{\mu} \right]_{s=1} \>
\end{eqnarray}
where
 $\delta z_1^{\mu}(t)$ was
assumed to vanish out of a small neighbourhood of a specific value of $t$.
Finally taking into account that
\begin{equation}
\delta u^{\mu}(1,t)= \delta z^{\mu}_1(t) \> , \qquad
\frac{\partial u^{\mu}_{\min}(1,t)}{\partial t} =
\dot{z}^{\mu}_1(t)
\end{equation}
 one obtains
\begin{eqnarray}
\delta (i \ln W_{q\overline{q}}^{\rm LR} ) = \sigma
\int_{t_{\rm i}}^{t_{\rm f}} dt \, \frac{1}{[{\cal S}]_{s=1}}
\left[ - \dot{z}^2_1 \left( \frac{\partial u_{\nu}^{\min}}{\partial s}
\right)_1 + \left( \frac{\partial u_{\mu}^{\min}}{\partial s}
\right)_1 \dot{z}_1^{\mu} \dot{z}_{1\nu} \right] \delta z_1^{\nu} =
\nonumber\\
= \frac{1}{2} \sigma \int_{t_{\rm i}}^{t_{\rm f}} dt \,
(dz^{\mu}_1 \delta z^{\nu}_1 - dz^{\nu}_1 \delta z^{\mu}_1)
\left[ \left( \frac{\partial u_{\mu}^{\min}}
{\partial s} \right)_1 \dot{z}_{1\nu} - \left(
\frac{\partial u_{\nu}^{\min}}{\partial s} \right)_1
\dot{z}_{1\mu} \right] \times
\nonumber\\
\times \left\{ - \dot{z}_1^2 \left( \frac{\partial u^{\min}}
{\partial s} \right)_1^2 + \left[ \dot{z}_1 \cdot \left(
\frac{\partial u^{\min}}{\partial s} \right)_1 \right]^2
\right\}^{-\frac{1}{2}}
\end{eqnarray}
and so Eq. (2.41).

\begin{figure}
\caption{The two types of configurations of the three-quark system.}
\end{figure}

\begin{figure}
\caption{Generalized Wilson loop for the quark-antiquark system.}
\end{figure}

\begin{figure}
\caption{The analogous of the Wilson loop for the three-quark system.}
\end{figure}

\end{document}